\documentclass[prd,showpacs,amsmath,amssymb]{revtex4}
\usepackage{dcolumn}
\usepackage{bm}
\usepackage{epsfig,graphicx}
\newcommand{\be}{\begin{equation}}
\newcommand{\ee}{\end{equation}}
\newcommand{\bea}{\begin{eqnarray}}
\newcommand{\eea}{\end{eqnarray}}
\newcommand{\bean}{\begin{eqnarray*}}
\newcommand{\eean}{\end{eqnarray*}}

\newcommand{\gapproxeq}{\lower
.7ex\hbox{$\;\stackrel{\textstyle >}{\sim}\;$}}
\newcommand{\lapproxeq}{\lower
.7ex\hbox{$\;\stackrel{\textstyle <}{\sim}\;$}}
\begin{document}

\bibliographystyle{unsrt}

\title{\bf Revisit the radiative decays of $J/\psi$ and $\psi'\to \gamma\eta_c (\gamma\eta_c^\prime)$ }

\author{Gang Li$^1$ and Qiang Zhao$^{2,3}$}

\affiliation{ 1) Department of Physics, Qufu Normal University,
Qufu, 273165, P.R. China}

\affiliation{ 2) Institute of High Energy Physics,
       Chinese Academy of Sciences, Beijing 100049, P.R. China
}
\affiliation{3) Theoretical Physics Center for Science Facilities,
Chinese Academy of Sciences, Beijing 100049, P.R. China}

\date{\today}

\begin{abstract}
With the new measurements of $J/\psi$ and $\psi^\prime\to
\gamma\eta_c \ (\gamma\eta_c^\prime)$ from CLEO and BES-III
Collaboration, we re-investigate the intermediate meson loop (IML)
contributions to these radiative decays in association with the
quark model M1 transitions in an effective Lagrangian approach. It
shows that the ``unquenched" effects due to the intermediate hadron
loops can be better quantified by the new data for $J/\psi\to
\gamma\eta_c$. Although the IML contributions are relatively small
in $J/\psi\to \gamma\eta_c$, they play a crucial role in
$\psi^\prime\to \gamma\eta_c \ (\gamma\eta_c^\prime)$. A prediction
for the IML contributions to $\psi(3770)\to \gamma\eta_c \
(\gamma\eta_c^\prime)$ is made. Such ``unquenched" effects allow us
to reach a coherent description of those three radiative
transitions, and gain some insights into the underlying dynamics.

\end{abstract}



\maketitle
\section{Introduction}

There has been a long-standing puzzle on the radiative transition
rates of $J/\psi$ and $\psi^\prime\to \gamma\eta_c \
(\gamma\eta_c^\prime)$. On the one hand, the nonrelativistic
potential model (NR model)  including color Coulomb plus linear
scalar potential, and spin-spin, spin-orbit interactions, has made
great successes in the description of the charmonium spectrum based
on the constituent degrees of freedom~\cite{Appelquist:1974yr}.
Nevertheless, a relativised version by Godfrey and Isgur (GI model)
also offers a reasonably good description of the hadron spectra and
transition matrix elements for quarkonia made of either light or
heavy $q\bar{q}$~\cite{Godfrey:1985xj,Barnes:2005pb}. On the other
hand, both NR and GI model have predicted relatively larger
branching ratios for $J/\psi$ and $\psi^\prime\to \gamma\eta_c \
(\gamma\eta_c^\prime)$. In particular, the predicted partial decay
width for $\psi^\prime\to \gamma\eta_c$ was nearly one order of
magnitude larger than the experimental data~\cite{Nakamura:2010zzi}.
In contrast with the success of the quark model in the description
of various properties of charmonium spectrum, such discrepancies
seem not to be trivial and have initiated a lot of theoretical
interests in the literature~\cite{Eichten:2004uh,Eichten:1974af,Eichten:1975ag,Eichten:1978tg,Eichten:1979ms,Ebert:2002pp,Eichten:2007qx,Brambilla:2005zw,Brambilla:2004wf}.

Recently the lattice QCD (LQCD) calculations of the charmonium
radiative transitions were
reported~\cite{Dudek:2006ej,Dudek:2009kk}. As shown by
Ref.~\cite{Dudek:2006ej}, in the ``quenched" approximation the
magnetic dipole (M1) transition of $J/\psi\to \gamma\eta_c$ was
consistent with the new experimental data from CLEO
collaboration~\cite{:2008fb}, although one notices that the lattice
value does not overlap with the experimental uncertainties. For
$\psi^\prime\to \gamma\eta_c^\prime$, the LQCD uncertainties are
even larger than the experimental ones, from which one cannot
conclude that ``unquenched" effects would not play a role here.

In Ref.~\cite{Brambilla:2005zw}, the M1 transition of $J/\psi\to
\gamma\eta_c$ was investigated in the framework of nonrelativistic
effective field theory. By assuming the ground state charmonium to
be a weakly coupled system, the authors obtained the radiative decay
width $\Gamma(J/\psi\to \gamma\eta_c)=(1.5\pm 1.0)$ keV to the
correction of ${\cal O}(v_c^2/m_c^2)$ with a large uncertainty due
to high-order corrections. For $\psi^\prime\to \gamma\eta_c
(\gamma\eta_c^\prime)$, the weakly-coupled-$c\bar{c}$ assumption
cannot be applied. Nevertheless, the mass of $\psi^\prime$ is close
to the open $D\bar{D}$ threshold, which may have non-negligible
effects on the constituent quark potential.

In Ref.~\cite{Li:2007xr}, we proposed to consider the intermediate
meson loop (IML) corrections as an ``unquenched" mechanism in the
charmonium energy region. Such a mechanism turns out to be important
for exclusive transitions especially when the mass of the initial
state is close to the open channel
threshold\cite{Li:1996yn,Cheng:2004ru,Anisovich:1995zu,Zhao:2006dv,Zhao:2006cx,Wu:2007jh,Liu:2006dq,Zhao:2005ip,Li:2007au,Zhang:2009kr,Liu:2009vv,Liu:2010um,Wang:2010iq,Guo:2010zk,Guo:2010ak}.
An evidential observation of the IML contributions should be the
$\psi(3770)$ non-$D\bar{D}$ decays. Since $\psi(3770)$ is close to
the $D\bar{D}$ threshold, the $D\bar{D}$ rescatterings into light
hadrons would be an essential process contributing to its
non-$D\bar{D}$ branching ratios~\cite{Zhang:2009kr}. In
Ref.~\cite{Li:2007xr}, it was shown that the $J/\psi$ exclusive
decays would experience relatively smaller open charm effects than
the $\psi^\prime$ since the latter is much closer to the $D\bar{D}$
threshold. In another word, the IML would have more important impact
on the $\psi^\prime$ decays, while the $J/\psi$ suffers less.

During the past two years, important progresses have been achieved
in the experimental measurements of these radiative transitions. The
CLEO Collaboration reported the branching ratios $BR(J/\psi\to
\gamma\eta_c)=(1.98\pm 0.09\pm 0.30)\%$ and
$BR(\psi^\prime\to\gamma\eta_c)=(4.32\pm 0.16\pm 0.60)\times
10^{-3}$~\cite{:2008fb}, while a search for the decay of
$\psi^\prime\to\gamma\eta_c^\prime$ only led to
$BR(\psi^\prime\to\gamma\eta_c^\prime)< 7.6\times
10^{-4}$~\cite{:2009vg}. In fact, the CLEO data for
$\psi^\prime\to\gamma\eta_c$ have been greatly weighted in the
PDG2010 averages, i.e. $BR(J/\psi\to\gamma\eta_c)=(1.7\pm 0.4)\%$
and $BR(\psi^\prime\to\gamma\eta_c)=(3.4\pm 0.5)\times
10^{-3}$~\cite{Nakamura:2010zzi}. Very recently, BES-III
Collaboration also reported the branching ratio
$BR(\psi^\prime\to\gamma\eta_c^\prime)=(4.7\pm 0.9_{stat}\pm
3.0_{sys})\times 10^{-4}$~\cite{besiii-hadron2011} as the first
measurement of this decay channel.

The above progresses in both theory and experiment thus prompt us to
revisit this problem based on the following considerations and
improvements of our calculation: i) As shown by the CLEO
data~\cite{:2008fb} and LQCD unquenched
calculations~\cite{Dudek:2006ej}, there leaves only small
corrections from the ``unquenched" effects in
$J/\psi\to\gamma\eta_c$. This will impose strong constraints on our
model parameters. As mentioned earlier that the IML should have
larger impact on the $\psi^\prime$ decays, the interest is to
investigate how important the intermediate $D$ meson loops would be
in the $\psi^\prime$ channel. ii) In Ref.~\cite{Li:2007xr}, the
$D^{*0}\to D^0\gamma$ coupling was fixed by the experimental upper
limit for the $D^{*0}$ total width. This should have overestimated
the loop contributions involving $D^{*0}D^0\gamma$ vertex. In this
work, we adopt a more realistic coupling value in the calculation.
iii) It was noticed in Refs.~\cite{Zhang:2009gy,Zhang:2010zv} that
the $D^*\bar{D}^*(D)$ loop were rather important. Thus, we include
all the $S$-wave $D$ mesons in the loop calculations to estimate the
leading contributions from the IML. iv) We do not include the
contact terms in this work. Such contributions, though turned out to
be negligibly small, were considered in Ref.~\cite{Li:2007xr}. The
contact terms were induced by EM minimal substitution at the $VVP$
vertex. We shall discuss later that the contact terms would be
eliminated in the ELA in order to avoid unphysical contributions. v)
The IML transitions also provide a mechanism to evade the quark
model selection rule at leading order for the $M1$ transition
between a $D$-wave and $S$-wave states. Thus, the IML contributions
in $\psi(3770)$ should be investigated.

This paper is organized as below. In Sec. II we present the
framework of the effective Lagrangian approach (ELA) for the IML.
Section III is devoted to numerical results and discussions. A brief
summary is given in the last section.

\section{Effective Lagrangian Approach for the IML}

The IML transitions, or known as final state interactions (FSI),
have been one of the important non-perturbative transition
mechanisms in many
processes~\cite{Li:1996yn,Cheng:2004ru,Anisovich:1995zu,Zhao:2006dv,Zhao:2006cx,Wu:2007jh,Liu:2006dq,Zhao:2005ip,Li:2007au,Zhang:2009kr,Liu:2009vv,Liu:2010um,Wang:2010iq,Guo:2010zk,Guo:2010ak}.
In the energy region of charmonium masses, with more and more data
from Belle, BaBar, CLEO and BES-III, it is widely recognized that
intermediate hadron loops may be closely related to a lot of
non-perturbative phenomena observed in experiment, e.g. apparent
OZI-rule
violations~\cite{Wu:2007jh,Liu:2006dq,Cheng:2004ru,Anisovich:1995zu,Zhao:2005ip,Li:2007au,Zhang:2009kr,Liu:2009vv,Liu:2010um,Wang:2010iq,Guo:2010zk,Guo:2010ak},
sizeable non-$D\bar D$ decay branching ratios for
$\psi(3770)$~\cite{Zhang:2009kr}, and the helicity selection rule
violations in charmonium
decays~\cite{Liu:2009vv,Liu:2010um,Wang:2010iq}. The IML transitions
play a role as ``unquenching" the simple $c \bar c$ picture in the
quark model. It can be easily understood since we know that the
charm quark is somehow not heavy enough. The failure of the heavy
quark approximation will then manifest itself in some exclusive
transitions, such as the problems investigated in this work.

Before proceed to the IML formulation, we recall the M1 transition
amplitude based on the constituent scenario as the following:
\begin{eqnarray}\label{eq:M1}
\Gamma_{M1}(n^{2S+1}L_J\to n'^{2S'+1}L'_{J'}+\gamma)= \frac
{4(2J'+1)} {3(2J+1)} \delta_{LL'} \delta_{S,S'\pm 1} \frac
{e_c^2\alpha} {m_c^2} |\langle \psi_f|\psi_i\rangle|^2 E_\gamma^3
\frac {E_f} {M_i} ,
\end{eqnarray}
where $n(n')$, $S(S')$, $L(L')$, $J(J')$,
$|\psi_i\rangle(|\psi_f\rangle)$  are the initial (final) state main
quantum number, spin, orbital angular momentum, total angular
momentum and spatial wavefunctions, respectively. $E_\gamma$ and
$E_f$ denote the final state photon and meson energy, respectively,
while $M_i$ is the initial $c\bar c$ meson mass.  The above equation
can be regarded as the ``quenched" contributions. The GI model M1
radiative rates do not incorporate the phase factor $E_f/M_i$, while
include a recoil factor $j_0(kr/2)$. In this work, we will quote the
results of Ref.~\cite{Barnes:2005pb}.

In the VVP transition, all mechanisms that contribute to this
transition will appear as corrections to a single anti-symmetric
tensor coupling. We derive the effective $V\gamma P$ couplings via
the M1 transition of Eq.~(\ref{eq:M1}) as follows,
\begin{eqnarray}
{{\cal M}_{fi}(M1)}=g_{V\gamma P} \varepsilon_{\alpha\beta\mu\nu}
p_i^\alpha\varepsilon_i^\beta p_\gamma^\mu \varepsilon_\gamma^\nu \,
,
\end{eqnarray}
where $p_i$ and $p_\gamma$ are four-vector momentum of the initial
meson and final state photon, respectively, and $\varepsilon_i$ and
$\varepsilon_\gamma$ are the corresponding polarization vectors.

\begin{figure}[tb]
\begin{center}
\vglue-0mm
\includegraphics[width=0.7\textwidth]{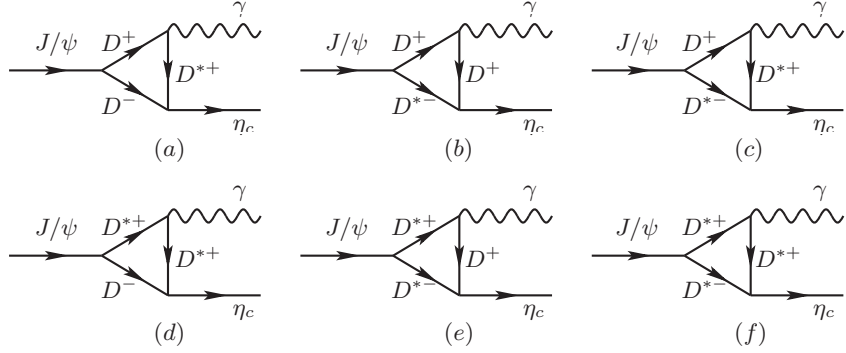}
\vglue-0mm \caption{Schematic picture for the decay of $J/\psi\to
\gamma \eta_c$ via (a) $D^+ D^-(D^{*+})$, (b) $D^+ D^{*-}(D^{+})$,
(c) $D^+ D^{*-}(D^{*+})$, (d) $D^{*+} D^{-}(D^{*+})$, (e) $D^{*+}
D^{*-}(D^{+})$, (f) $D^{*+} D^{*-}(D^{*+})$ intermediate charmed
meson loops. Similar diagrams are for strange charmed mesons. For
neutral charmed mesons, only (a), (c) and (e) can contribute.
Similar pictures occur in $\psi'\to \gamma \eta_c$ and $\gamma
\eta_c'$.\label{fig:Tri}}
\end{center}
\end{figure}

\begin{figure}[tb]
\begin{center}
\vglue-0mm
\includegraphics[width=0.7\textwidth]{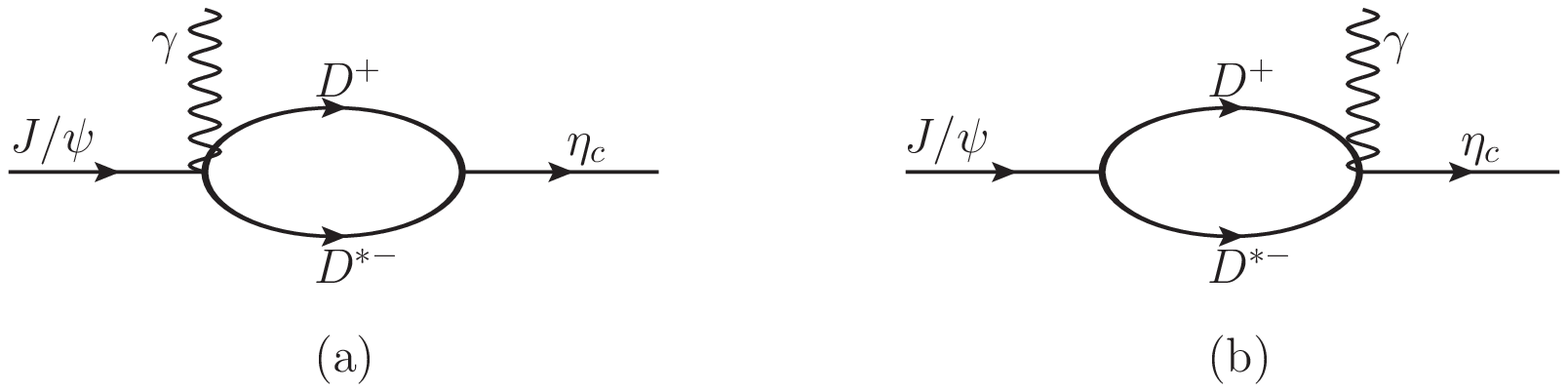}
\includegraphics[width=0.7\textwidth]{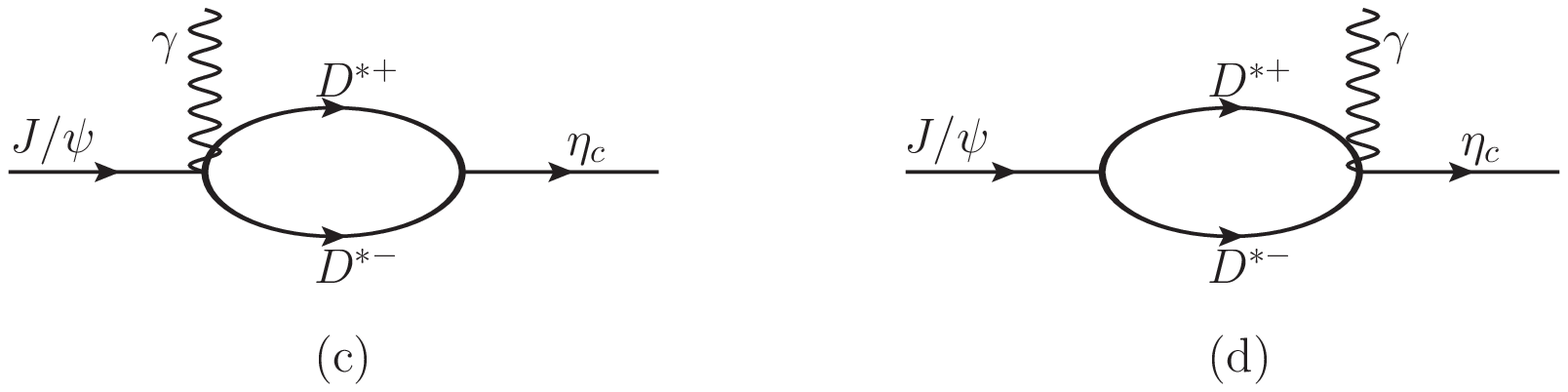}\\
\vglue-0mm \caption{The contact diagrams considered in $J/\psi\to
\gamma \eta_c$.  Similar diagrams are also considered in $\psi'\to
\gamma \eta_c(\eta_c').$ \label{fig:contact}}
\end{center}
\end{figure}

The IML transitions can be schematically illustrated by the triangle
(Fig.~\ref{fig:Tri}) and contact processes (Fig.~\ref{fig:contact}).
The coupling between an $S$-wave charmonium and charmed mesons is
given by the effective Lagrangian based on heavy quark
symmetry~\cite{Colangelo:2003sa,Casalbuoni:1996pg},
\begin{equation}
\mathcal{L}_2=i g_2 Tr[R_{c\bar{c}} \bar{H}_{2i}\gamma^\mu
{\stackrel{\leftrightarrow}{\partial}}_\mu \bar{H}_{1i}] + H.c.,
\end{equation}
where the $S$-wave vector and pseudoscalar charmonium states are
expressed as
\begin{equation}
R_{c\bar{c}}=\left( \frac{1+ \rlap{/}{v} }{2} \right)\left(\psi^\mu
\gamma_\mu -\eta_c \gamma_5 \right )\left( \frac{1- \rlap{/}{v} }{2}
\right).
\end{equation}
The charmed and anti-charmed meson triplet are
\begin{eqnarray}
H_{1i}&=&\left( \frac{1+ \rlap{/}{v} }{2} \right)[
\mathcal{D}_i^{*\mu}
\gamma_\mu -\mathcal{D}_i\gamma_5], \\
H_{2i}&=& [\bar{\mathcal{D}}_i^{*\mu} \gamma_\mu
-\bar{\mathcal{D}}_i\gamma_5]\left( \frac{1- \rlap{/}{v} }{2}
\right),
\end{eqnarray}
where $\mathcal{D}$ and $\mathcal{D}^*$ are pseudoscalar
($(D^{0},D^{+},D_s^{+})$) and vector charmed mesons
($(D^{*0},D^{*+},D_s^{*+})$), respectively.

Explicitly, the Lagrangian for the $S$-wave charmonium ($J/\psi$,
$\psi'$ and $\eta_c$) couplings to $D$ and $D^*$ becomes
\begin{eqnarray}\label{LS}
\mathcal{L}_S &=& ig_{\psi \mathcal{D}^* \mathcal{D}^*} (-\psi^\mu
\mathcal{D}^{*\nu}\overleftrightarrow{\partial}_\mu \mathcal{D}_\nu^{*\dagger}+
\psi^\mu \mathcal{D}^{*\nu}\partial_\nu\mathcal{D}^{*\dagger}_{\mu} -
\psi_\mu\partial_\nu \mathcal{D}^{*\mu} \mathcal{D}^{*\nu\dagger})\nonumber\\
&&+ ig_{\psi \mathcal{D}\mathcal{D}}\psi_\mu(\partial^\mu
\mathcal{D}\mathcal{D}^{\dagger}-\mathcal{D}\partial^\mu
\mathcal{D}^{\dagger}) + g_{\psi
\mathcal{D}^*\mathcal{D}}\varepsilon^{\mu\nu\alpha\beta}\partial_\mu
\psi_\nu(\mathcal{D}^*_\alpha\overleftrightarrow{\partial}_\beta
\mathcal{D}^{\dagger}
- \mathcal{D}\overleftrightarrow{\partial}_\beta\mathcal{D}^{*\dagger}_\alpha) \nonumber \\
&& + g_{\eta_c D^* D} D^{*\mu}(\partial_\mu \eta_c D-\eta_c
\partial_\mu D)  + ig_{\eta_c D^* D^*} \varepsilon^{\mu\nu\alpha\beta}
\partial_\mu D^\ast_\nu D^{*\dagger}_\alpha \partial_\beta\eta_c \ .
\end{eqnarray}

The effective Lagrangians for the electromagnetic (EM) interaction
vertices of $\gamma DD$, $\gamma D^*D^*$ and $\gamma DD^*$ are
~\cite{Dong:2009uf}
\begin{eqnarray}\label{ELA-photon}
{\cal L}_{DD\gamma}(x)&=& ie A_\mu(x)D^-(x)
\overleftrightarrow{\partial}^\mu D^+(x)  + ieA_\mu D_s^-(x)
\overleftrightarrow{\partial}^\mu D_s^+(x)
\\
{\cal L}_{D^*D^*\gamma}(x)&=&
ie A_\mu(x) \, \biggl\{ g^{\alpha\beta} \,
D^{*-}_\alpha \overleftrightarrow{\partial}^\mu D^{*+}_\beta(x)
+ g^{\mu\beta} \,
D^{*-}_\alpha(x) \partial^\alpha D^{*+}_\beta(x)
- g^{\mu\alpha} \,
\partial^\beta D^{*-}_\alpha(x) D^{*+}_\beta(x) \biggr\} \nonumber \\
&+&ie A_\mu(x) \, \biggl\{ g^{\alpha\beta} \,
D^{*-}_{s\alpha} \overleftrightarrow{\partial}^\mu D^{*+}_{s\beta}(x)
+ g^{\mu\beta} \,
D^{*-}_{s\alpha}(x) \partial^\alpha D^{*+}_{s\beta}(x)
- g^{\mu\alpha} \,
\partial^\beta D^{*-}_{s\alpha}(x) D^{*+}_{s\beta}(x) \biggr\}
\,, \\
{\cal L}_{D^*D\gamma}(x)&=&
\frac{e}{4} \epsilon^{\mu\nu\alpha\beta}F_{\mu\nu}(x)
\biggl\{  g_{D^{*-}D^+\gamma} D^{*-}_{\alpha\beta}(x)D^+(x)
        + g_{D^{*0}D^0\gamma} \bar{D}^{*0}_{\alpha\beta}(x)D^0(x)\nonumber \\
        &+&g_{D_s^{*-}D_s^+\gamma} D^{*-}_{s\alpha\beta}(x)D_s^+(x)
\biggr\} + \ {\rm H.c.} \,,
\end{eqnarray}
where $A \overleftrightarrow{\partial}_{\mu} B \equiv  A
\partial_{\mu} B -
\partial_{\mu} A B$,
$F_{\mu\nu}\equiv \partial_{\mu}A_{\nu}-\partial_{\nu}A_{\mu}$, and
$M_{\mu\nu}\equiv \partial_{\mu}M_{\nu}-\partial_{\nu}M_{\mu}$. The
EM interaction for neutral $D$ meson ($D^0D^0\gamma$ and
$D^{*0}D^{*0}\gamma$) do not exist. The coupling constants appearing
in the effective Lagrangians will be determined later.

The loop transition amplitudes in  Fig.~\ref{fig:Tri} can be
expressed in a general form in the ELA as follows:
 \begin{eqnarray}
 M_{fi}=\int \frac {d^4 p_2} {(2\pi)^4} \sum_{D^\ast \ \mbox{pol.}}
 \frac {T_1T_2T_3} {a_1 a_2 a_3}{\cal F}(m_2,p_2^2)
 \end{eqnarray}
where $T_i \ (i=1,2,3)$ are the vertex functions; $a_i \equiv
p_i^2-m_i^2 \ (i=1,2,3)$ are the denominators of the intermediate
meson  propagators. We adopt the typical dipole form factor in the
calculation, i.e.
\begin{equation}\label{ELA-form-factor}
{\cal F}(m_2,p_2^2)\equiv \left(\frac
{\Lambda^2-m_2^2} {\Lambda^2-p_2^2}\right)^2,
\end{equation}
 where
$\Lambda\equiv m_2+\alpha\Lambda_{QCD}$ and the QCD energy scale,
$\Lambda_{QCD} = 220$ MeV. This form factor will take care of the
non-local effects of the vertex functions and kill the loop
divergence in the integrals. The value of parameter $\alpha$ is
commonly at the order of unity.

In Fig.~\ref{fig:Tri}, the triangle diagrams for charged
intermediate meson loops are illustrated. For the neutral ones, only
those corresponding to diagrams (a), (c) and (e) can contribute.
Diagrams (b), (d) and (f) have no contributions due to the vanishing
$D^0 {\bar D}^0\gamma$ and $D^{0*}{\bar D}^{*0}\gamma $ couplings.
Note that it is easy to check that all the diagrams in
Fig.~\ref{fig:Tri} satisfy gauge invariance individually. The
explicit transition amplitudes for those triangle loops are given as
follows:
\begin{eqnarray}\nonumber
{\cal M}_T^{(a)}&=&(i^3)\int \frac{d^4p_2} {(2\pi)^4}[g_{\psi
DD}\varepsilon_\psi^\rho (p_{1\rho}-p_{3\rho})] [-g_{\eta_c
D^*D}(p_{f\sigma}+p_{3\sigma})][-eg_{D^*D\gamma}\varepsilon_{\alpha\beta\mu\nu}
p_\gamma^\alpha  \varepsilon_\gamma^{*\beta} p_2^\mu]\\
&\times &\frac {i} {p_1^2-m_1^2} \frac {i(-g^{\nu\sigma}+p_2^\nu
p_2^\sigma/m_2^2)} {p_2^2-m_2^2} \frac {i} {p_3^2-m_3^2}{\cal
F}(m_2,p_2^2) \ , \\
{\cal M}_T^{(b)}&=&(i^3)\int \frac{d^4p_2} {(2\pi)^4}[g_{\psi
D^*D}\varepsilon_{\alpha\beta\mu\nu}p_\psi^\alpha\varepsilon_\psi^\beta
(p_3^\nu-p_1^\nu)] [-g_{\eta_c D^*D} (p_{f\rho}+p_{2\rho})][ e
\varepsilon_\gamma^{*\sigma}(p_{1\sigma}+p_{2\sigma})] \nonumber \\
&\times &\frac {i} {p_1^2-m_1^2} \frac {i} {p_2^2-m_2^2} \frac
{i(-g^{\mu\rho}+p_3^\mu p_3^\rho/m_3^2)} {p_3^2-m_3^2}{\cal
F}(m_2,p_2^2) \ , \\
{\cal M}_T^{(c)}&=&(i^3)\int \frac{d^4p_2} {(2\pi)^4}[g_{\psi
D^*D}\varepsilon_{\alpha\beta\mu\nu}p_\psi^\alpha\varepsilon_\psi^\beta
(p_3^\nu-p_1^\nu)] [g_{\eta_c
D^*D^*}\varepsilon_{\rho\sigma\xi\tau}p_2^\rho p_f^\tau][-eg_{ D^*D
\gamma}
\varepsilon_{\theta\phi\kappa\lambda}p_\gamma^\theta\varepsilon_\gamma^{*\phi}
p_2^\kappa ] \nonumber \\
&\times &\frac {i} {p_1^2-m_1^2} \frac
{i(-g^{\sigma\lambda}+p_2^\sigma p_2^\lambda/m_2^2)} {p_2^2-m_2^2}
\frac {i(-g^{\mu\xi}+p_3^\mu p_3^\xi/m_3^2)} {p_3^2-m_3^2}{\cal
F}(m_2,p_2^2) \ , \\
{\cal M}_T^{(d)}&=&(i^3)\int \frac{d^4p_2} {(2\pi)^4}[-g_{\psi
D^*D}\varepsilon_{\alpha\beta\mu\nu}p_\psi^\alpha\varepsilon_\psi^\beta(p_{3}^\nu-p_{1}^\nu)][-g_{\eta_c
D^*D}(p_{f\theta}+p_{3\theta})] \nonumber \\
&\times & [e\varepsilon_{\gamma}^{*\rho}
(-g_{\sigma\tau}(p_{1\rho}+p_{2\rho})
+g_{\rho\tau} p_{1\sigma}-g_{\rho\sigma} p_{2\tau})] \nonumber\\
&\times &\frac {i(-g^{\mu\tau}+p_1^\mu p_1^\tau/m_1^2)}
{p_1^2-m_1^2} \frac {i(-g^{\theta\sigma}+p_2^\theta
p_2^\sigma/m_2^2)} {p_2^2-m_2^2} \frac {i} {p_3^2-m_3^2}{\cal
F}(m_2,p_2^2) \ , \\
{\cal M}_T^{(e)}&=&(i^3)\int \frac{d^4p_2}
{(2\pi)^4}\varepsilon_\psi^\mu[g_{\psi
D^*D^*}((p_{1\mu}-p_{3\mu})g_{\alpha\beta}+g_{\beta\mu}
p_{3\alpha}-g_{\alpha\mu} p_{1\beta})][-g_{\eta_c
D^*D}(p_{f\nu}+p_{3\nu})] \nonumber \\
&\times &
[-eg_{D^*D\gamma}\varepsilon_{\rho\sigma\xi\tau}p_\gamma^\rho\varepsilon_\gamma^{*\sigma}
p_{1}^\xi]\frac {i(-g^{\alpha\tau}+p_1^\alpha p_1^\tau/m_1^2)}
{p_1^2-m_1^2} \frac {i}
{p_2^2-m_2^2} \frac {i(-g^{\beta\nu}+p_3^\beta p_3^\nu/m_3^2)} {p_3^2-m_3^2}{\cal F}(m_2,p_2^2) \ , \\
{\cal M}_T^{(f)}&=&(i^3)\int \frac{d^4p_2}
{(2\pi)^4}\varepsilon_\psi^\mu[g_{\psi
D^*D^*}((p_{1\mu}-p_{3\mu})g_{\alpha\beta}+g_{\beta\mu}
p_{3\alpha}-g_{\alpha\mu} p_{1\beta})] [g_{\eta_c
D^*D^*}\varepsilon_{\rho\sigma\xi\tau} p_{2}^\rho p_{f}^\tau]
\nonumber \\
&\times & [-e\varepsilon_{\gamma}^{*\theta}(g_{\phi\kappa}
(p_{2\theta}+p_{1\theta})+g_{\kappa\theta} p_{1\phi} -
g_{\phi\theta} p_{2\kappa})]\frac {i(-g^{\alpha\kappa}+p_1^\alpha
p_1^\kappa/m_1^2)} {p_1^2-m_1^2} \nonumber \\ &\times & \frac
{i(-g^{\sigma\phi}+p_2^\sigma p_2^\phi/m_2^2)} {p_2^2-m_2^2} \frac
{i(-g^{\beta\xi}+p_3^\beta p_3^\xi/m_3^2)} {p_3^2-m_3^2}{\cal
F}(m_2,p_2^2) \ .
\end{eqnarray}

The contact diagrams of Figs.~\ref{fig:contact}(a), (b), (c) and (d)
arise from gauging the strong $J/\psi D^*\bar{D}$ and $\eta_c
D^*\bar{D}$ $J/\psi D^*\bar{D}^*$ and $\eta_c D^*\bar{D}^*$
interaction Lagrangians by the minimal substitution $\partial^\mu\to
\partial^\mu + ie A^\mu$. One can also easily check that gauge
invariance is guaranteed for the contact diagrams in
Fig.~\ref{fig:contact}. However, these contact diagrams can be
neglected in the calculation based on the following detailed
examinations. The processes in Figs.~\ref{fig:contact} can be
classified into two categories. The first one includes diagrams (a)
and (d), where the contact vertices induced by the EM minimal
substitution violate gauge invariance. We discard these
contributions based on the empirical argument that an initial
massive vector meson decaying into $\gamma VP$ via the contact
interaction actually violates gauge invariance, thus is forbidden.
The second category includes diagrams (b) and (c) in
Figs.~\ref{fig:contact}. Although these contact vertices keep gauge
invariant, they do not have contributions due to the vanishing loop
integrals as shown in Ref.~\cite{Li:2007xr}. In brief, we argue that
the contact diagrams would not contribute to the transition matrix
elements in the ELA. Therefore, we can concentrate on the triangle
diagrams in the following calculations.

In principle, we should include all the possible triangle meson
loops in the calculation. In reality, the breakdown of the local
quark-hadron duality allows us to pick up the leading contributions
as a reasonable approximation~\cite{Lipkin:1986bi,Lipkin:1986av}.
Also intermediate states involving flavor changes turn out to be
strongly suppressed. One reason is because of the large virtualities
involved in the light meson loops. The other is because of the OZI
rule suppressions. So we will only consider the charmed meson loops
as the leading contributions in this work. It should be noted that
the IML transitions can naturally evade the quark model selection
rule as a dynamical mechanism.

\section{Numerical Results}

To proceed the numerical results, we first clarify the following
points:

(i) In the heavy quark limit, the couplings for charmonium-charmed
mesons in Eq.~(\ref{LS}) can be related to the parameter $g_2$
defined in Ref.~\cite{Colangelo:2003sa},
\begin{eqnarray}
\label{eq:couplings_ELA} g_{\psi \mathcal{D}\mathcal{D}} & = &
2g_2\sqrt{m_\psi} m_\mathcal{D}, \,\,\,\,\,\,\, g_{\psi
\mathcal{D}^*\mathcal{D}^*} =  -2g_2\sqrt{m_\psi} m_{\mathcal{D}^*},
\,\,\,\,\,\,\, g_{\psi \mathcal{D}^\ast \mathcal{D}} =
2g_2\sqrt{\frac {m_\psi m_\mathcal{D}} {m_{\mathcal{D}^*}}},
\end{eqnarray}
where $g_2\equiv \sqrt{m_\psi}/(2m_\mathcal{D} f_\psi)$ and the
$f_\psi=405$ MeV is the $J/\psi$ decay constant.

(ii) The ratios of the couplings constants $g_{\psi' DD}$ and
$g_{\psi'D^*D^*}$  to $g_{J/\psi DD}$ and $g_{J/\psi D^*D^*}$ are
fixed as
\begin{eqnarray}
\frac {g_{\psi'DD}} {g_{J/\psi DD}} = \frac {g_{\psi'D^*D^*}}
{g_{J/\psi D^*D^*}} = 0.9 \ .
\end{eqnarray}
This ratio has uncertainties as adopted in the
literature~\cite{Dong:2009uf}. A recent study of the $e^+ e^-\to
D\bar{D}$ cross section lineshape~\cite{Zhang:2009gy} suggests that
$g_{\psi'DD} \simeq g_{J/\psi DD}$, and we will discuss later that
to coherently account for the partial widths of $J/\psi$ and
$\psi^\prime\to \gamma\eta_c \ (\gamma\eta_c^\prime)$ would impose a
strong constraint on the ratio of $g_{\psi'DD}/ g_{J/\psi DD}$.

(iii) The radiative couplings for $D^*\to D\gamma$ can be determined
by the partial widths from experimental measurement, i.e.
\begin{eqnarray}
\Gamma(D^*\to D\gamma)=\frac{\alpha}{3} g_{D^*D\gamma}^2|P_\gamma|^3
\ .
\end{eqnarray}
The partial width $\Gamma(D^{*+}\to D^+ \gamma)$ has been precisely
measured, i.e. $\Gamma(D^{*+}\to D^+ \gamma)= 1.54$
keV~\cite{Nakamura:2010zzi}. This allows us to extract
$g_{D^{*+}D^+\gamma} = -0.5 \ {\mbox GeV}^{-1}$. For $D^{*0}\to D^0
\gamma$, the branching ratio is measured by
experiment~\cite{Nakamura:2010zzi}. However, the total width of
$D^{*0}$ has not been well determined. This will bring some
uncertainties to the estimate of the radiative coupling. However,
this quantity can be related to the $g_{D^{*+}D^+\gamma}$ in the
constituent quark model, which gives $g_{D^{*0}D^0\gamma}  = +2.0 \
{\mbox GeV}^{-1}$. This is also a value obtained in different
approaches. For the coupling $g_{D_s^*D_s\gamma}$, the value
$(-0.3\pm 0.1) \ {\mbox GeV}^{-1}$ from QCD sum rules
(QSR)~\cite{Zhu:1996qy} is adopted. We note that their relative
signs are consistent with each other in the framework of
LQCD~\cite{Becirevic:2009xp}, QSR~\cite{Zhu:1996qy}, and constituent
quark model.

(iv) Another two undetermined parameters in our model are the
cut-off parameter $\alpha$ in the form factor and the relative phase
$\delta$ between the ``quenched" (i.e. quark model M1 transition
amplitude) and ``unquenched" (i.e. IML transition) amplitude. Taking
the advantage of the anti-symmetric tensor coupling for $VVP$, we
can always parametrize the total amplitude as
\begin{eqnarray}
{\cal M}_{fi}=[g_{V\gamma P}+
\tilde{g}_{tri}e^{i\delta}]\varepsilon_{\alpha\beta\mu\nu}p_\gamma^\alpha
\varepsilon_\gamma^\beta p_i^\mu \varepsilon_i^\nu \ ,
\end{eqnarray}
where $g_{V\gamma P}$ is fixed to be a positive real number by the
``quenched" quark model M1 transition amplitude, and
$\tilde{g}_{tri}$ is extracted from the triangle loop diagrams. We
simply take $\delta=0$ or $\pi$ in the calculation, since the decay
threshold of the intermediate mesons are above the initial meson (
$J/\psi$ and $\psi'$) masses, the absorptive part of the loop
integrals does not contribute as a consequence. In fact, because the
``quenched" quark model M1 transition amplitude has overestimated
the experimental data, it determines $\delta=\pi$ in the
calculation.

Note that in the ``quenched" quark model scenario the spin-flipping
M1 transition amplitude for $J/\psi\to \gamma\eta_c$ shares the same
sign as that for $\psi^\prime\to \gamma\eta_c^\prime$ since the
latter pair states are just the corresponding radial excitations of
the former ones. In this sense, the same phase angle $\delta=\pi$
for $\psi^\prime\to \gamma\eta_c^\prime$ seems to be reasonable. It
is interesting to observe that this phase relation is also respected
in $\psi^\prime\to\gamma\eta_c$.

With the destructive phase $\delta=\pi$, the parameter
$\alpha=0.98\pm 0.27$ is determined  by combining the GI model
result with the IML to reproduce the experimental partial width
$\Gamma_{exp}(J/\psi \to \gamma \eta_c)=(1.58\pm 0.37)$ keV
\cite{Nakamura:2010zzi}. The range of $\alpha$ is given by the
experimental error bars. We then apply the same set of $\alpha$ and
$\delta$ to predict the partial width of $\psi'\to \gamma \eta_c
(\gamma\eta_c')$, and ``unquenched" effects in $\psi(3770)\to
\gamma\eta_c$ and $\gamma\eta_c^\prime$.

\begin{table}
\begin{tabular}{|c|c|c|c|c|c|}
 \hline
 Initial meson                 & $J/\psi(1^3S_1)$   &\multicolumn{2}{c|}      {$\psi'(2^3S_1)$}       & \multicolumn{2}{c|} {$\psi{''}(1^3 D_1$)}  \\
 \hline
 Final meson                   & $\eta_c(1^1S_0)$   & $\eta_c'(2^1S_0)$   &  $\eta_c(1^1S_0)$         & $\eta_c'(2^1S_0)$   &  $\eta_c(1^1S_0)$\\
 \hline
 $\Gamma^{NR}_{M1}$ (keV)      & 2.9                & 0.21                      &  9.7                & ---& ---      \\
 \hline
 $\Gamma^{GI}_{M1}$ (keV)      & 2.4                & 0.17                      &  9.6                & ---& ---     \\
 \hline
 $\Gamma_{IML}$     (keV)      & $0.08_{-0.06}^{+0.13}$   & $0.02_{-0.01}^{+0.02}$        & $2.78_{-1.96}^{+5.73}$        & $1.82_{-1.19}^{+1.95}$  & $17.14_{-12.03}^{+22.93}$\\
 \hline
 $\Gamma_{all}$     (keV)      & $1.58^{+0.37}_{-0.37}$    & $0.08^{+0.03}_{-0.03}$             & $2.05^{+2.65}_{-1.75}$    & $1.82_{-1.19}^{+1.95}$  & $17.14_{-12.03}^{+22.93}$     \\
 \hline
 $\Gamma_{exp}$     (keV)      & $1.58\pm 0.37$~\cite{Nakamura:2010zzi}     & $0.143\pm 0.027\pm 0.092$~\cite{besiii-hadron2011}                      & $0.97\pm 0.14$~\cite{:2008fb}      & ---& ---     \\
 \hline
 $\Gamma_{LQCD}$     (keV)      & $2.51\pm 0.08$     & ---                      & $0.4\pm 0.8$      & ---& $10\pm 11$     \\
 \hline
\end{tabular}
\caption{Radiative partial decay widths given by different processes
are listed: $\Gamma^{NR}_{M1}$ and $\Gamma^{GI}_{M1}$ are the M1
transitions in the NR and GI model,
respectively~\cite{Barnes:2005pb}. $\Gamma_{IML}$ are contributions
from the IML transitions (Fig.~\ref{fig:Tri}), and $\Gamma_{all}$
are coherent results including the M1 transition amplitude of the GI
model and IML transitions. The experimental data are from
PDG~\cite{Nakamura:2010zzi}, BES-III~\cite{besiii-hadron2011}, and
CLEO~\cite{:2008fb}. The LQCD results are also included as a
reference~\cite{Dudek:2009kk}. } \protect\label{tab-1}
\end{table}

\begin{figure}[tb]
\begin{center}
\vglue-0mm
\includegraphics[width=0.7\textwidth]{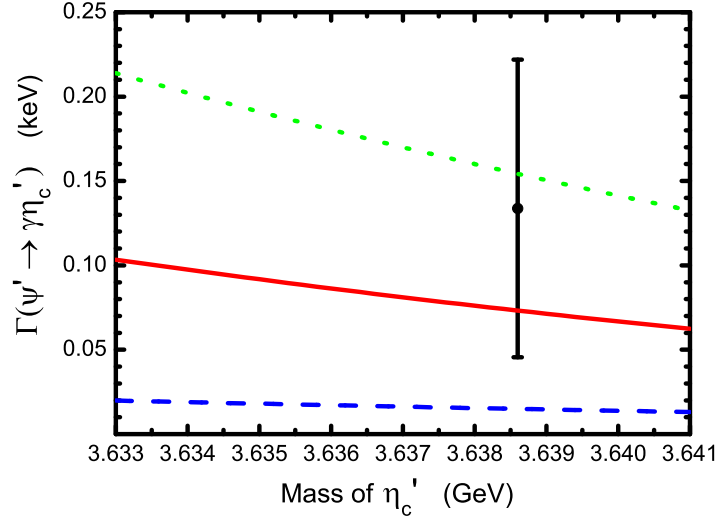}
\vglue-0mm \caption{Partial widths of $\psi'\to \gamma \eta_c'$ in
terms of the mass of $\eta_c'$ with $\alpha=0.98$. The dashed line
is the exclusive IML contributions, and the dotted line is the
exclusive GI model result. The solid line is the combined result of
both. The datum point is from the BES-III
Collaboration~\cite{besiii-hadron2011}. \label{fig:3}}
\end{center}
\end{figure}

In Tab.~\ref{tab-1}, the calculated M1 transition branching ratios
are listed to compare with the GI and NR quark model results. The
partial width of $J/\psi\to \gamma\eta_c$ is an input to fix
$\alpha$.  The LQCD calculations are also included as a reference.
In comparison with the previous estimate in Ref.~\cite{Li:2007xr},
the requirement of smaller ``unquenched" effects leads to smaller
IML contributions in all these three decay channels.

For $\psi^\prime\to \gamma\eta_c \ (\gamma\eta_c^\prime)$, we learn
the following points: (i) The ``unquenched" contributions from the
IML still play a role of destructive interferences with the GI
amplitude to bring down the M1 transition amplitudes. The calculated
partial decay widths are consistent with the BES-III preliminary
result of $\Gamma(\psi^\prime\to
\gamma\eta_c^\prime)$~\cite{besiii-hadron2011}, and CLEO measurement
of $\Gamma(\psi^\prime\to \gamma\eta_c)$~\cite{:2008fb}. (ii) One
notices that the uncertainty with the form factor parameter $\alpha$
extracted in $J/\psi\to \gamma \eta_c$ still causes large
uncertainties in the estimate of the IML contributions in the
$\psi^\prime$ decays. This shows a sensitivity correlation of the
IML contributions for the $n ^3S_1 \to n'  {^1S_0}$ M1 transitions.
One also notices that the partial width $\Gamma(\psi^\prime\to
\gamma\eta_c^\prime)$ has relatively smaller uncertainties than
$\Gamma(\psi^\prime\to \gamma\eta_c)$. This is self-consistent since
the former transition does not violate the selection rule of
Eq.~(\ref{eq:M1}) and the ``quenched"  quark model leading order
transition is still dominant.

For $\psi(3770)\to \gamma\eta_c \ (\gamma\eta_c^\prime)$, if
$\psi(3770)$ is a pure $D$-wave state, the M1 transition will be
forbidden by the selection rule of Eq.~(\ref{eq:M1}). However, due
to the nonvanishing photon energy in the decay, higher multipoles
beyond the leading one would contribute. In a harmonic oscillator
basis, the nonvanishing transition amplitude of $\psi(3770)\to\gamma
\eta_c (\eta_c^\prime)$ is the same order as that of
$\psi^\prime\to\gamma\eta_c \ (\eta_c^\prime)$. Since a quantitative
estimate of the quark model amplitude will depend on the details of
model constructions, we only concentrate here the IML mechanism that
present the ``unquenched" contributions. As listed in
Tab.~\ref{tab-1}, the IML transitions predict $\Gamma(\psi(3770)\to
\gamma\eta_c)=(17.14_{-12.03}^{+22.93})$ keV and
$\Gamma(\psi(3770)\to \gamma\eta_c^\prime)=(1.82_{-1.19}^{+1.95})$
keV, which are in a reasonable order of magnitude, although
uncertainties appear to be significant. This is similar to
$\psi^\prime\to\gamma\eta_c$, where sensitivity of the partial
widths to the range of $\alpha$ values is obvious. Interestingly, it
shows that the IML contributions are the same order as the LQCD
results~\cite{Dudek:2009kk}. This implies that interferences between
the ``quenched" and ``unquenched" amplitudes should be important for
the $\psi(3770)$ radiative decays. As a consequence, the radiative
transition of $\psi(3770)$ could become either abnormally strong or
significantly small in comparison with potential quark model
expectations. Experimental measurement of these radiative
transitions would be helpful for providing further constraint on the
IML contributions.

It is interesting to note that the present experimental data for
$J/\psi$ and $\psi'\to \gamma\eta_c \ (\gamma\eta_c')$ would tightly
stretch the parameter space for the form factor parameter $\alpha$
and coupling $g_{\psi'DD}$. Since these processes should share the
same form factor parameter $\alpha$, the main parameter difference
is the coupling between $J/\psi (\psi')$ to $D^{(*)}\bar{D}^{(*)}$
for which the analysis of Ref.~\cite{Zhang:2009gy} suggests that a
smaller value for $g_{\psi'DD}$ should be applied.

For $\psi^\prime\to\gamma\eta_c^\prime$, the mass of $\eta_c^\prime$
may also cause uncertainties to the extracted IML contribution. With
the fixed $\alpha=0.98$, we investigate the sensitivities of the IML
contributions to the $\eta_c^\prime$ mass. In Fig.~\ref{fig:3}, we
plot the exclusive partial width  $\Gamma(\psi^\prime\to \gamma
\eta_c^\prime)$ from the IML in terms of the mass of $\eta_c'$
within a range of 3.633$\sim$ 3.641 GeV~\cite{Nakamura:2010zzi}. It
shows that within the PDG mass range, the IML contributions are
rather stable. The partial width decreases in term of the increasing
$m_{\eta_c'}$ due to the decreasing phase space in the decay
transition.

\section{Summary}

We revisited the hadronic meson loop contributions to  the $J/\psi$
and $\psi'$ radiative decays into $\gamma\eta_c$ or $\gamma\eta_c'$
in the ELA. In the framework of an improved effective Lagrangian
approach, the IML transitions provide ``unquenched" corrections to
the leading couplings extracted from potential quark models due to
the unique Lorentz structure of $VVP$ interactions. In comparison
with the NR and GI model, the IML contributions intend to cancel the
quark model ``quenched" amplitudes. Apart from those more elaborate
treatments for the meson loop calculations, we have applied an
experimental constraint on the strong coupling of $g_{\psi'DD}$
based on the analysis of the cross section lineshape of $e^+e^-\to
D\bar{D}$~\cite{Zhang:2009gy}, where a relatively smaller value of
$g_{\psi'DD}$ was favored. It is interesting to see that the IML
effects in $J/\psi\to \gamma\eta_c$ are much smaller than that in
$\psi'\to\gamma\eta_c \ (\gamma\eta_c')$. Note that the pure M1
contribution in $\psi'\to \gamma \eta_c$ is about one order of
magnitude larger than the experimental data, such a significant
discrepancy implies the necessity of ``unquenching" the quark model
scenario especially when the transitions are close to open
thresholds.

For $\psi(3770)\to \gamma\eta_c \ (\gamma\eta_c')$, we predict quite
significant corrections from the IML, which are the same order of
magnitude as the LQCD ``quenched" result~\cite{Dudek:2009kk}. This
is an interesting issue related to the $\psi(3770)$ non-$D\bar{D}$
decays. The BES-III Collaboration recently scanned over the
$\psi(3770)$ mass region. It will be possible to measure the
radiative decays of $\psi(3770)\to \gamma\eta_c \ (\gamma\eta_c')$,
and further clarify the role played by the IML effects as an
``unquenched" mechanism for the $c\bar{c}$ quark model scenario.

\section*{Acknowledgements}
This work is supported, in part, by the National Natural Science
Foundation of China (Grant No. 11035006 and No. 10947007), the
Chinese Academy of Sciences (KJCX2-EW-N01), the Ministry of Science
and Technology of China (2009CB825200), the Natural Science
Foundation of Shandong Province under Grant No. ZR2010AM011 and the
Scientific Research Starting Foundation of Qufu Normal University.

\end{document}